%
\let\useblackboard=\iftrue 
%
%
\newfam\black 
\input harvmac.tex 
\input tables.tex 
%
\input epsf.tex 
\ifx\epsfbox\UnDeFiNeD\message{(NO epsf.tex, FIGURES WILL BE 
IGNORED)} 
\def\figin#1{\vskip2in}
\else\message{(FIGURES WILL BE INCLUDED)}\def\figin#1{#1}\fi 
\def\ifig#1#2#3{\xdef#1{fig.~\the\figno} 
\midinsert{\centerline{\figin{#3}}%
\smallskip\centerline{\vbox{\baselineskip12pt 
\advance\hsize by -1truein\noindent{\bf Fig.~\the\figno:} #2}} 
\bigskip}\endinsert\global\advance\figno by1} 
\noblackbox 
\def\Title#1#2{\rightline{#1} 
\ifx\answ\bigans\nopagenumbers\pageno0\vskip1in%
\baselineskip 15pt plus 1pt minus 1pt 
\else
\def\listrefs{\footatend\vskip 
1in\immediate\closeout\rfile\writestoppt 
\baselineskip=14pt\centerline{{\bf 
References}}\bigskip{\frenchspacing%
\parindent=20pt\escapechar=` \input 
refs.tmp\vfill\eject}\nonfrenchspacing} 
\pageno1\vskip.8in\fi \centerline{\titlefont #2}\vskip .5in} 
  
scaled\magstep3 
  
scaled\magstep3 
  
scaled\magstep3 
  
scaled\magstep3 
  
scaled\magstep3 
\ifx\answ\bigans\def\tcbreak#1{}\else\def\tcbreak#1{\cr&{#1}}\fi 

\useblackboard 
\message{If you do not have msbm (blackboard bold) fonts,} 
\message{change the option at the top of the tex file.}

\font\blackboard=msbm10 scaled \magstep1 
\font\blackboards=msbm7 
\font\blackboardss=msbm5 
\textfont\black=\blackboard 
\scriptfont\black=\blackboards 
\scriptscriptfont\black=\blackboardss 
\def\Bbb#1{{\fam\black\relax#1}} 
\else 
\def\Bbb{\bf} 
\fi 
%

%
\def\yboxit#1#2{\vbox{\hrule height #1 \hbox{\vrule width #1 
\vbox{#2}\vrule width #1 }\hrule height #1 }} 
\def\fillbox#1{\hbox to #1{\vbox to #1{\vfil}\hfil}} 
\def\ybox{{\lower 1.3pt \yboxit{0.4pt}{\fillbox{8pt}}\hskip-0.2pt}} 

\def\r{\right} 
\def\comments#1{}

\def\QC{\Bbb{C}}

\def\QZ{\Bbb{Z}}
\def\QP{\Bbb{P}} 
\def\p{\partial}

\def\Tr{{{\rm Tr\  }}}

\def\CN{{\cal N}}

\def\II{\relax{I\kern-.07em I}}

\def\inbar{\,\vrule height1.5ex width.4pt depth0pt} 
\def\IZ{\relax\ifmmode\mathchoice 
{\hbox{\cmss Z\kern-.4em Z}}{\hbox{\cmss Z\kern-.4em Z}} 
{\lower.9pt\hbox{\cmsss Z\kern-.4em Z}} 
{\lower1.2pt\hbox{\cmsss Z\kern-.4em Z}}\else{\cmss Z\kern-.4em 
Z}\fi} 
\def\IB{\relax{\rm I\kern-.18em B}} 
\def\IC{{\relax\hbox{$\inbar\kern-.3em{\rm C}$}}} 
\def\ID{\relax{\rm I\kern-.18em D}} 
\def\IE{\relax{\rm I\kern-.18em E}} 
\def\IF{\relax{\rm I\kern-.18em F}} 
\def\IG{\relax\hbox{$\inbar\kern-.3em{\rm G}$}} 
\def\IGa{\relax\hbox{${\rm I}\kern-.18em\Gamma$}} 
\def\IH{\relax{\rm I\kern-.18em H}} 
\def\IK{\relax{\rm I\kern-.18em K}} 
\def\IP{\relax{\rm I\kern-.18em P}} 
\def\pp{{\relax{=\kern-.42em |\kern+.2em}}} 

\def\p{\partial}

\font\cmss=cmss10 \font\cmsss=cmss10 at 7pt 
\def\IR{\relax{\rm I\kern-.18em R}}

\def\Tr{{\rm Tr\ }}


%
%

%
 
\Title{ \vbox{\baselineskip12pt\hbox{hep-th/9812025} 
\hbox{TUW-98-24} 
}} 
{\vbox{ 
\centerline{A Family of $\CN=1$ $SU(N)^k$ Theories from}
\vskip4mm
\centerline{Branes at Singularities}}} 

\centerline{Esperanza Lopez$^{\natural}$} 
\medskip 
\centerline{$^\natural$ Institut f\"ur theoretische Physik, TU-Wien} 
\centerline{Wiedner Hauptstra{\ss}e 8-10} 
\centerline{A-1040 Wien, Austria}  
\centerline{\tt elopez@tph44.tuwien.ac.at} 
\vskip1cm 
 
\centerline{\bf{Abstract}} 
We obtain $\CN=1$ $SU(N)^k$ gauge theories with bifundamental 
matter and a quartic superpotential as the low energy theory 
on D3-branes at singular points. These theories generalize that 
on D3-branes at a conifold point, studied recently by Klebanov 
and Witten. For $k=3$ the defining 
equation of the singular point is that of an isolated $D_4$ 
singularity. For $k>3$ we obtain a family of multimodular singularities.  
The considered $SU(N)^k$ theories flow in the infrared to a
non-trivial fixed point. We analyze the $AdS/CFT$ 
correspondence for our examples.

\vfill 
\Date{\vbox{\hbox{\sl December, 1998}}} 
 
\lref\p{J. Polchinski, ``Dirichlet-Branes and Ramond-Ramond Charges'',
Phys.Rev.Lett. 75 (1995) 4724, hep-th/9510017.}
\lref\kw{I. R. Klebanov and E. Witten, ``Superconformal Field Theory on 
Threebranes at a Calabi-Yau Singularity'', hep-th/9807080.}
\lref\arnold{V. I. Arnold, S. M. Gusein-Zade and A. N. Varchenko,
``Singularities of Differentiable Maps'', vol.I, Birkh\"auser, 1985.}
\lref\mope{D. R. Morrison and M. R. Plesser, ``Non-Spherical 
Horizons, I'', hep-th/9810201.}
\lref\u{A. M. Uranga, ``Brane Configurations for Branes at Conifolds'',
hep-th/9811004.}
\lref\dm{M. R. Douglas and G. Moore, ``D-branes, Quivers, and ALE 
Instantons'', hep-th/9603167.}
\lref\dgm{M. Douglas, B. Greene and D. Morrison, ``Orbifold Resolution
by D-branes, hep-th/9704151.}
\lref\lnv{A. Lawrence, N. Nekrasov and C. Vafa, ``On Conformal Theories 
in Four Dimensions'', Nucl.Phys. B533 (1998) 199, hep-th/9803015.}
\lref\kk{}
\lref\m{J. M. Maldacena, ``The Large N Limit of Superconformal Field 
Theories and Supergravity'', Adv.Theor.Math.Phys. 2 (1998) 231,
hep-th/9711200.}
\lref\k{A. Kehagias, ``New Type IIB Vacua and their F-Theory 
Interpretation'', Phys.Lett. B435 (1998) 337, hep-th/9805131.}
\lref\aj{O. Aharony, A. Fayyazuddin and J. Maldacena, ``The Large N 
Limit of ${\cal N} =2,1 $ Field Theories from Threebranes in F-theory,
J.High Energy Phys. 07 (1998) 013, hep-th/9806159.}  
\lref\cone{}
\lref\sei{N. Seiberg, ``Electric-Magnetic Duality in Supersymmetric 
Non-Abelian Gauge Theories'', Nucl.Phys. B435 (1995) 129, hep-th/9411149.}
\lref\w{E. Witten, ``Baryons And Branes In Anti de Sitter Space'', 
J.High Energy Phys. 9807 (1998) 006, hep-th/9805112.}
\lref\gk{S. S. Gubser and I. R. Klebanov, ``Baryons and Domain Walls in 
an N = 1 Superconformal Gauge Theory'', hep-th/9808075.}
\lref\conifold{}
\lref\milnor{J. Milnor, ``Singular Points of Complex Hypersurfaces'',
Ann. of Math. Studies, 61.}
\lref\st{R. G. Leigh and M. J. Strassler, ``Exactly Marginal Operators 
and Duality in Four Dimensional N=1 Supersymmetric Gauge Theory'', 
Nucl. Phys. B447 (1995) 95, hep-th/9503121.}
\lref\gkp{S. S. Gubser, I. R. Klebanov and A. M. Polyakov, ``Gauge 
Theory Correlators from Non-Critical String Theory'', Phys.Lett. B428 
(1998) 105, hep-th/9802109.}
\lref\holw{E. Witten, ``Anti De Sitter Space And Holography'', 
Adv.Theor.Math.Phys. 2 (1998) 253, hep-th/9802150.}
\lref\radu{K. Oh and R. Tatar, ``Three Dimensional SCFT from M2 Branes 
at Conifold Singularities'', hep-th/9810244.}
\lref\lu{L. E. Ibanez, R. Rabadan and A. M. Uranga, ``Anomalous U(1)'s 
in Type I and Type IIB D=4, N=1 string vacua'', hep-th/9808139.}
\lref\vafacon{K. Hori, H. Ooguri and C. Vafa, ``Non-Abelian Conifold 
Transitions and N=4 Dualities in Three Dimensions'', Nucl.Phys. B504 
(1997) 147, hep-th/9705220.}
\lref\kasi{S. Kachru and E. Silverstein, ``4d Conformal Field Theories 
and Strings on Orbifolds'', Phys.Rev.Lett. 80 (1998) 4855, hep-th/9802183.}
\lref\muki{K. Dasgupta and S. Mukhi, ``Brane Constructions, Conifolds 
and M-Theory'', hep-th/9811139.} 
\lref\r{M. Reid, ``Young Person's Guide to Canonical Singularities'',
Algebraic Geometry, Bowdoin, 1985, Proc. Sympos. Pure Math., vol. 46,
part 1, Amer. Math. Soc., Providence, RI, 1987, pp. 345.}
\lref\rr{M. Reid, ``Canonical 3-folds'', Journe\'ees de G\'eom\'etrie
Alg\'ebriqe d'Angers (A. Beauville, ed.) Sitjhoff \& Noordhoof, 1980,
pp. 273.}
\lref\taylor{M. A. Luty and W. Taylor IV, ``Varieties of vacua in 
classical supersymmetric gauge theories'', Phys.Rev. D53 (1996) 3399,
hep-th/9506098.}
\lref\gns{S. Gubser, N. Nekrasov and S. Shatashvili, ``Generalized 
Conifolds and 4d N=1 SCFT'', hep-th/9811230.}
\lref\gu{S. S. Gubser, ``Einstein manifolds and conformal field theories'', 
hep-th/9807164.}
\lref\afbh{B. S. Acharya, J. M. Figueroa-O'Farrill, C. M. Hull, and B. Spence, ``Branes at conical singularities and holography'', hep-th/9808014.}
\lref\wtwo{E. Witten, ``Solutions Of Four-Dimensional Field Theories Via M 
Theory'', Nucl.Phys. B500 (1997) 3, hep-th/9703166.}

\newsec{Introduction}

Dirichlet-branes \p\ have proved to be an extremely useful tool 
for the study of gauge theories. The low-energy theory of $N$ 
coincident D-branes placed at a regular point of the transversal 
space is maximally supersymmetric $U(N)$ Yang-Mills. More general 
theories with less supersymmetry can be obtained by placing D-branes 
at singular points of the transversal space. Particular examples
that have been extensively studied in the literature are D-branes 
at orbifold points \dm\ \dgm\ \lnv.
 
In \m\ Maldacena proposed a very interesting duality between
large $N$ gauge theories and type IIB or M-theory in a 
background given by the near horizon geometry of black-branes.
The near horizon geometry of 3-branes at a regular point of
the transverse space is $AdS_5 \times S^5$. The conjectured 
duality proposes that $\CN=4$ $SU(N)$
Yang-Mills theory is dual to type IIB string theory on 
$AdS_5 \times S^5$, where $S^5$ bears $N$ units of five-form 
flux. There has been a big effort in extending this duality 
to the case of 3-branes sitting at a singular point of the
transverse space. In that case $S^5$ is substituted by a
five-dimensional manifold $H$, which describes 
the angular part of the singular space. When the singular
space has an orbifold description, $H$ is given by $S^5/\Gamma$ where
$\Gamma$ is the discrete orbifold group \kasi. For $\Gamma \subset SU(2)$, 
$SU(3)$ or $SU(4)$ the dual gauge theories have $\CN=2,1$
or $0$ conformal symmetry respectively.   

Klebanov and Witten considered D3-branes at a conifold point in 
\kw. The associated gauge theory is $\CN=1$ $SU(N) \times SU(N)$ 
(plus a free $U(1)$)
with bifundamental matter multiplets and a quartic 
superpotential. The $AdS/CFT$ correspondence for this case has been 
analyzed in great detail \kw\ \gu\ \gk. Branes at spaces with 
more general singularities have been recently considered in \mope\ 
\radu\ \u, where the associated gauge theories were also derived. 
Our aim in this paper is to find a class of singular spaces such 
that the world-volume theory on D3-branes at the singular 
point generalizes that of \kw\ to the case of $k$ factor groups. 
In order to achieve this goal, we will follow a somehow opposite 
approach to that used in \kw\ \mope\ \radu\ \u. Instead of deriving 
the field 
theory once the singular space is known, we will use the field 
theory data to construct the singular three-fold. Section 2 will
be devoted to studying the case $k=3$. We will find that the
associated singular space describes an isolated $D_4$-singularity 
of a three-fold. 

Our $SU(N)^k$ theories flow in the infrared to a non-trivial 
fixed point. In section 3 we will analyze the $AdS/CFT$ 
correspondence for the case $k=3$. We will see that making certain 
assumptions about the topology of the $D_4$-singularity, we obtain 
a consistent $AdS/CFT$ correspondence. In particular, we will 
compare moduli spaces of gauge and string theory, global symmetries
of the gauge theory with gauge symmetries of string dual and 
the spectrum of baryonic operators with branes wrapped
on homology cycles of $H$ \gk\ \w. In section 4 we generalize
our results to $k>3$. 

After this paper was completed, we received reference \gns, where
the same problem is treated.

\newsec{Branes at Singularities}

The gauge theory on $N$ parallel D3-branes at a conifold
point is $\CN=1$ $U(N) \times U(N)$ with four 
chiral multiplets $A_i$, $B_j$, $i,j=1,2$, transforming in the 
$({\bf N},{\bf \bar N})$ and $({\bf \bar N},{\bf N})$ representations 
of the gauge group respectively \kw. 
We wish to find a singular space such that the gauge
theory on D3-branes at the singular point is the 
generalization of the previous one to the case of three factor 
groups, i.e. $\CN=1$ $U(N) \times U(N) \times U(N)$ with 
chiral matter fields 
\eqn\fields{       
\eqalign{       
A =  ({\bf N},{\bf \bar N},1), \;\;\;\; B= (1,{\bf N},{\bf \bar N}),  
\;\;\;\; & C=({\bf \bar N},1,{\bf N}), \cr   
{\tilde A}=({\bf \bar N},{\bf N},1), \;\;\;\; {\tilde B}=(1,{\bf \bar N},
{\bf N}), \;\;\;\; & 
{\tilde C}=({\bf N},1,{\bf \bar N}). }}
The superpotential of the gauge theory will be determined by the criterion 
that a branch of the moduli space can be interpreted as positions 
of D3-branes. 

It is convenient to consider first the case $N=1$. This should correspond 
to a single D3-brane in our searched for singular space. The D-term 
equations are
\eqn\Dterms{
\eqalign{
| A |^2 + | {\tilde C} |^2 - |{\tilde A} |^2 - | C |^2 &
= \xi_1, \cr
| B |^2 + | {\tilde A} |^2 - | {\tilde B}|^2 - | A |^2 &
= \xi_2, \cr
| C |^2 + | {\tilde B} |^2 - | {\tilde C}|^2 - | B |^2 &
= \xi_3. }}
The real numbers $\xi_i$ are Fayet-Iliopoulos parameters. Since the sum 
of the lhs's in \Dterms\ is zero, the parameters $\xi_i$ must satisfy 
$\sum \xi_i=0$ in order to allow for a supersymmetric vacuum. 
The fact that only two of the three D-term equations are linearly
independent implies that there is a combination of $U(1)$'s under
which all the fields are uncharged. This $U(1)$ field will be thus
free and decouple. We will denote by $Q$ the space of solutions to 
\Dterms\ quotiented by the gauge transformations generated by the two 
non-trivial $U(1)$'s. The space $Q$ has complex dimension four.

We can describe $Q$ directly in terms of gauge invariant 
quantities \taylor. A minimal set of gauge invariant quantities is
\eqn\CYfields{
\eqalign{
& x_1 =  A {\tilde A}, \;\;\;\;  x_2= B {\tilde B}, \;\;\;\; x_3= 
C {\tilde C}, \cr & z=  ABC, \;\;\; w= {\tilde A}{\tilde B} 
{\tilde C} . }
}
These variables are not independent, they are subject to the 
constraint
\eqn\CY{ x_1 x_2 x_3 = z w.
}
This equation defines a hypersurface in $\QC^5$. 
The space \CY\ is singular along the following codimension three
subspaces: {\it i)} $x_1=x_2=y=z=0$; {\it ii)} $x_1=x_3=y=z=0$;  
{\it iii)} $x_2=x_3=y=z=0$. The three lines of singularities 
intersect at the origin $x_i=y=z=0$. When $\xi_i=0$, $Q$ coincides 
with the four-fold \CY. Values $\xi_i \neq 0$ (for some 
$i$) correspond to (partial) resolutions of \CY\ in which a singular 
subspace is substituted by $\Sigma \times \QC$, where $\Sigma$ denotes 
a two-sphere.
  
The moduli space of the gauge theory is the subspace of $Q$ determined 
by the F-term equations. Let us consider the quartic superpotential
\eqn\spone{
W= (a_1 x_1 + a_2 x_2 + a_3 x_3)^2,
}
where $a_i$ are complex parameters. The F-term equations derived 
from $W$ reduce to a single relation expressible in terms of the 
variables $x_i$
\eqn\CYadd{
a_1 x_1 + a_2 x_2 + a_3 x_3=0.
}
The moduli space of the gauge theory is given by the intersection 
between $Q$ and the hyperplane \CYadd\ in $\QC^5$. This
defines a six-dimensional space, thus susceptible of being 
interpreted as the transversal space to a D3-brane.
\spone\ is the most general quartic superpotential with this
property. For $a_i \neq 0$ the
only singular point contained in the intersection between $Q$
and \CYadd\ is the origin. We can use the hyperplane equation 
to eliminate one of the $x_i$ variables in \CY. When $a_i \neq 0$,
we obtain 
\eqn\CYsix{
-\left( {a_1 \over a_2} x_1 +{a_3 \over a_2} x_3 \right) x_1 x_3 = zw.
}
We will denote this space by $K$. It is indeed only singular at the origin. 
By an obvious linear change of coordinates we can rewrite it as
\eqn\dfour{
x^3  + y^2 x = z w,}
which is the standard form of a $D_4$ singularity of a complex
three-fold \arnold. Since the parameters $a_i$ can be eliminated
by a coordinate change, they do not affect the complex structure
of $K$.

The superpotential \spone\ can be extended to that of an $U(N)^3$ theory
with matter content \fields\
\eqn\sp{
\eqalign{
W= & \; a_1^2 \; \Tr(A {\tilde A})^2 + a_2^2 \; \Tr(B {\tilde B})^2 + 
a_3^2 \; \Tr(C {\tilde C})^2 + 2 \; a_1 a_3 \; 
\Tr A {\tilde A} {\tilde C} C + \cr 
& 2 \; a_1 a_2 \; \Tr B {\tilde B} {\tilde A} A + 
2 \; a_2 a_3 \; \Tr C {\tilde C} {\tilde B} B.}
}
All the matter fields are uncharged under the diagonal combination
of the three $U(1)$ factors in $U(N)^3$, thus this $U(1)$ 
is free. The other two $U(1)$ fields have positive beta functions 
and are expected to decouple in the infrared limit \foot{For gauge 
theories on D3-branes at orbifold singularities the 
non-trivial $U(1)$ fields have been shown to be spontaneously broken 
\dm\ or anomalous \lu.}. Therefore from now
on we will actually work with an $SU(N)^3$ theory, instead of $U(N)^3$.
The F-term equations derived from \sp\ imply that 
\CYsix\ is still verified, but now by the matrix quantities
\eqn\op{
\eqalign{
x_1= A {\tilde A}, \;\;\;\;\;\;\;  & x_3= {\tilde C} C,  \cr
y= ABC,\;\;\;\;\;\; & z= {\tilde C}{\tilde B} {\tilde A} . }
}
Notice that all these quantities transform in the adjoint representation 
of the first $SU(N)$ group factor (plus a singlet). Of course, 
analogous relations can be derived involving operators which transform
in the adjoint of the second and the third $SU(N)$ factors.
A family of vacua solving the D- and F-term equations is 
given by matrices that are, in some basis, diagonal and such
that each entry satisfies \CYadd\ and \CYsix. Along these vacua 
the superpotential \sp, plus the Higgs mechanism, will
give masses to the non-diagonal excitations. This family 
of vacua reproduces $N$ copies of the space $K$. In order to obtain 
the moduli space of the theory we should quotient by the Weyl 
transformations of $SU(N)_i$. Since all the operators in \op\ 
transform in the adjoint representation of $SU(N)_1$, quotienting by 
Weyl transformations produces the moduli space $K^N/S_N$, where 
$S_N$ is the group of permutations of $N$ elements. This is precisely
the moduli space associated to positions of $N$ D3-branes in the
space $K$. 

In order that a certain singular complex manifold is a valid 
compactification space for string theory, the singularity must
be of a restricted type known as Gorenstein canonical singularity
\mope. This means that there exists a non-vanishing holomorphic
top form near the singularity that extends to a holomorphic
form on any smooth blow up of the singularity. The $D_4$ 
singularity is of this type \rr\ \r. Therefore it is consistent to 
interpret the $\CN=1$ $SU(N)^3$ gauge theory with matter content 
\fields\ and superpotential \sp\ as the low-energy theory of $N$ 
D3-branes at the singular space $K$. 

The $SU(N)^3$ theory has baryon operators $B_i=X_i^N$, ${\tilde B}_i=
{\tilde X}_i^N$ for $X_i=A,B,C$.
A non-zero expectation value for one of the baryon operators
higgses the original theory down to $SU(N) \times SU(N)$. 
Let us consider $\langle A \rangle 
\sim 1$ and substitute this in \sp. The non-zero vev induces
a mass term for ${\tilde A}$ and thus only $B, {\tilde B}$
and $C, {\tilde C}$ remain as light fields. Integrating out 
${\tilde A}$ we obtain the superpotential 
\eqn\spcone{
W= 2 a_2 a_3 \Tr ( C {\tilde C} {\tilde B} B - 
{\tilde C} C B {\tilde B}).
}
This is precisely the superpotential of the $SU(N)^2$ theory 
obtained on $N$ D3-branes at a conifold \kw. 
Along the baryonic branches, some of the $\xi_i$ parameters 
in \Dterms\ are non-zero. We have argued that non-zero
$\xi_i$ correspond to resolutions or partial resolutions of
the singular space transverse to the D3-branes. Thus, if
our $SU(N)^3$ theory lives on the world-volume 
of D3-branes at a $D_4$ singularity, it is necessary 
that this singularity can be partially resolved by a single blow 
up to a conifold. 

It is convenient to use the standard form \dfour\ of the $D_4$ 
singularity: $x^3 + y^2 x = z w$. We can partially resolve it 
by blowing up the space $(x,y,z,w)$ at $x=z=0$ as explained in 
\vafacon. The partially resolved surface is covered by open
sets $({\tilde x},y,z,w)=(x,y,{\tilde z},w)$, glued together
by the conditions ${\tilde x} {\tilde z}=1$ and $z=x {\tilde z}$.
The inverse image of our surface in the first set is $w=z^2 {\tilde x}^3 + 
y^2 {\tilde x}$, which is non-singular. In the second set we obtain
$x^2 + y^2 = {\tilde z} w$, which is the defining
equation of the conifold. The inverse image of the singular 
point $x=y=z=w=0$ is the $\QP_1$ parameterized by ${\tilde x}$
in the first patch and ${\tilde z}$ in the second patch.

The group of non-anomalous continuous global symmetries of the $SU(N)^3$ 
theory is $U(1)_R \times U(1)_1 \times U(1)_2 \times U(1)$, 
under which the fields transform as indicated in Table.1.
\thicksize=1pt
\vskip12pt
\begintable
\tstrut  | $\;\; U(1)_R \;\;$ | $\;\; U(1)_1 \;\;$| 
$\;\; U(1)_2 \;\;$ | $\;\; U(1) \;\;$  \crthick
$\; \; A \;\; $          | $1/2$ |  $\;\;\; 1$ | $\;\;\; -1$ | $\;\;\; 1$  \cr
$\;\; {\tilde A} \;\;$ | $1/2$ | $-1$ |  $\;\;\; 1$ | $-1$  \cr
$\;\; B \;\;$          | $1/2$ |  $\;\;\; 0$ |  $\;\;\; 1$ |  $\;\;\; 0$  \cr
$\;\; {\tilde B} \;\;$ | $1/2$ |  $\;\;\; 0$ | $-1$ |  $\;\;\; 0$  \cr
$\;\; C \;\;$          | $1/2$ | $-1$ | $0$ |  $\;\;\; 0$  \cr
$\;\; {\tilde C}  \;\;$ | $1/2$ |  $\;\;\; 1$ |  $\;\;\; 0$ |  $\;\;\; 0$  
\endtable
\vskip3mm
The $U(1)_1$ and $U(1)_2$ symmetries are associated with the 
two non-trivial $U(1)$ factors of the $U(N)^3$ gauge group
living on the D3-branes. These $U(1)$ gauge fields are 
expected to decouple in the infrared, but the transformations they
generate survive as global symmetries \mope. The superpotential 
\sp\ is also invariant under charge conjugation, $X_i \rightarrow 
{\tilde X}_i^t$, ${\tilde X}_i \rightarrow X_i^t$, with
$X_i=A,B,C$. Extra discrete symmetries occur for particular
values of the $a_i$ parameters in the superpotential. We will
discuss them in the next section.

An interesting limit of the theory is achieved when one of the parameters 
$a_i$ goes to zero. To be definite, we consider $a_3 \rightarrow 0$.
Sending $a_3$ to zero while keeping $a_1,a_2$ 
finite would imply that the fields $C$, ${\tilde C}$ do not appear
in the superpotential. This situation cannot represent a 
theory derived from parallel D3-branes. There are baryonic 
directions in which the theory gets higgsed down to $SU(N)$. 
This corresponds to move the $N$ D3-branes
away from the singular point while keeping them together.
The world-volume theory of D3-branes at a regular point of 
the transverse space must flow in the infrared to $\CN=4$ 
Yang-Mills. However if one of the $a_i$ is zero, the 
superpotential along the mentioned vacua does not reproduce 
that of an $\CN=4$ theory.
In order to avoid this problem we can perform a double
limit in which we send $a_3$ to zero and $a_1,a_2$ to 
infinity such that 
\eqn\limit{
a_1/a_2=a , \;\;\;\; a_2 a_3 =b
}
are kept finite. We obtain then the superpotential
\eqn\splimitint{
W=a_2^2 \Tr (B {\tilde B} + a \; {\tilde A} A)^2 + 2 b \;
\Tr (C {\tilde C} {\tilde B} B + a \; {\tilde C} C A {\tilde A} ).}
The first term on the right hand side becomes infinite in the limit 
that we are considering. Its form is such that it can be interpreted 
as a divergence due to integrating out a chiral field $\phi$ in the 
adjoint representation of the second $SU(N)$, with mass $m \sim 
a_2^{-2}$. Integrating in this field we get
\eqn\splimit{
W= c \Tr \phi (B{\tilde B} + a \; {\tilde A} A) + 2 b \; 
\Tr (C {\tilde C} {\tilde B} B + a \; {\tilde C} C A {\tilde A} ).
}
This theory was recently analyzed in \mope\ \u. There it was
proposed to be the world-volume theory on D3-branes at the singular
space
\eqn\spp{
x y^2 = z w.
}
We observe that performing the previous double limit in \CYsix\
we obtain, up to rescalings, this same space. It corresponds to
intersect $Q$ with a hyperplane containing one of the four-fold singular 
subspaces. We will denote it by $\tilde K$. It is singular at the 
codimension two subspace $y=z=w=0$.

The existence of a limit in which adjoint matter becomes massless
has the following origin. Our $\CN=1$ theory is related to an $\CN=2$ 
theory with the same gauge group and matter content. Notice that the 
fields \fields\ can be paired up to form three $\CN=2$ hypermultiplets. 
That $\CN=2$ gauge theory can be derived from D3-branes at a $\QZ_3$ 
orbifold, with $\QZ_3$ acting only on four of the six transversal 
coordinates \dm. The associated superpotential is
\eqn\spntwo{
W = \Tr \phi_1 (A {\tilde A} - {\tilde C} C) +
      \Tr \phi_2 (B {\tilde B} - {\tilde A}A) + 
     \Tr \phi_3 (C {\tilde C} - {\tilde B}B) ,
}
where $\phi_i$ are chiral multiplets transforming in the
adjoint of each gauge group factor. We can break ${\cal N}=2$ to 
${\cal N}=1$ by giving masses to the adjoint fields
\eqn\mass{
\Delta W= m_1 \Tr \phi_1^2 + m_2 \Tr \phi_2^2+ m_3 \Tr \phi_3^2.
}
Integrating out the adjoint fields we obtain the superpotential
\eqn\spnone{
W= -{1 \over 4 m_1} \Tr ( A {\tilde A} - {\tilde C} C)^2
 -{1 \over 4 m_2} \Tr ( B {\tilde B} - {\tilde A} A)^2
 -{1 \over 4 m_3} \Tr ( C {\tilde C} - {\tilde B} B)^2.
}
This superpotential is of the form \sp\ when $\sum m_i =0$.
The condition $\sum m_i=0$ implies that the mass perturbation 
\mass\ can be written as $m \Tr (\phi_1^2 - \phi_2^2) + 
m' \Tr (\phi_1^2 - \phi_3^2)$. Let us consider switching on 
$m'$ but keeping $m=0$. After integrating out $\phi_1$ and 
$\phi_3$, and redefining $\phi_2 \rightarrow \phi_2 - 
{1 \over 4 m'} (B {\tilde B} + {\tilde A}A)$, we obtain
\splimit\ with $c=-a=1$, $b= 1/{8 m'}$. The superpotential obtained 
by breaking $\CN=2$ to $\CN=1$ by adjoint masses contains less
free parameters than \sp\ or \splimit. The reason is that 
$\CN=2$ supersymmetry fixes the coupling between adjoint
fields and hypermultiplets. Once $\CN=2$ is broken new 
parameters can be introduced in the superpotential associated 
to these couplings, as it is clearly the case in \splimit.
The only restriction we impose on the parameters is that the 
space of solutions to the D- and F-terms equations can be interpreted 
as positions of D3-branes in a six-dimensional transverse space.
In the next section we will analyze which parameters survive in the
infrared limit of the theory.

\newsec{AdS Dual Theory}

In a generic situation, the world-volume theory on D3-branes couples 
to the bulk degrees of freedom of type IIB string theory. We can 
decouple the gravity degrees of freedom and still obtain an interacting 
world-volume theory by sending the string tension to infinity ($\alpha' 
\rightarrow 0$) while keeping the string coupling fixed.
This limit can also be performed in the supergravity metric
representing a collection of $N$ parallel black 3-branes. 
The resulting metric reproduces the near-horizon geometry 
of the black branes. In \m\ Maldacena conjectured that, for 
large $N$, the gauge theory living on the D3-branes is dual to 
type IIB string theory on a background given
by the near-horizon geometry of the black branes. In this 
section we want to apply this proposal to our example.  

The supergravity metric representing $N$ parallel 3-branes 
with flat world-volume placed at a certain point of a 
six-dimensional space is \k\ \aj\
\eqn\met{
{d s}^2 =  f(y)^{-1/2} {d x}^2 + f(y)^{1/2} {d y}^2,
}
where ${d x}^2$ is the flat four-dimensional Minkowski metric 
and ${d y}^2$ is the metric in the transversal space.  
The function $f(y)$ determines the expectation value of the
5-form field strength associated to the presence of the 
3-branes and satisfies
\eqn\fs{ 
\Delta f = -(2 \pi)^4 N {\delta^6(y -y_0) \over {\sqrt g}}, 
\;\;\;\; F_{0123i}=-{1 \over 4} \partial_i f^{-1},
}
with $\Delta$ denoting the Laplacian of the six-dimensional space, 
$y_0$ the point at which the $N$ 3-branes sit, and $0123$ being the 
directions along the world-volume of the 3-branes. 

We are interested in the case that the transverse space is $K$, given by 
\CYsix, and the 3-branes are placed at its singular point. Especially 
relevant is the case in which the metric on the transverse space, close 
to a singular point, can be written as \kw\ \afbh\ \mope\
\eqn\metrick{
{d y}^2= r^2 \left( {dr^2 \over r^2} + d \Omega_H^2 + O(r/ \alpha'^{1/2})
\right), }
where $r$ measures distances with respect to the singular point.
$H$ is a five-dimensional space given by points at a distance 
$r=1$ from the singular point, also denoted horizon manifold of the 
singularity \mope. 
$d \Omega_H^2$ is a dimensionless metric on $H$, independent of $r$ and 
$\alpha'$. Expression \metrick\ means that close to the singular point, 
the transverse space can be approximated by a cone over $H$. 
When \metrick\ holds, the metric \met-\fs\ turns in the decoupling of 
gravity limit into that of $AdS_5 \times H$ \m\ \kw. 
We will assume that $K$ admits a Ricci-flat metric that close to the 
singular point verifies \metrick. 

For the case of zero superpotential each sector of the $SU(N)^3$ gauge 
theory provides an ${\cal N} =1$ $SU(N)$ theory with $2N$ fundamental 
hypermultiplets. Although this theory has non-zero beta function, it is 
expected to flow in the infrared to a non-trivial fixed point \sei. 
The superpotential \sp\ is non-renormalizable as a peturbation of the free 
field theory, but it is a marginal peturbation of the superconformal 
theory \kw.
We follow \kw\ in interpreting the $AdS/CFT$ correspondence for our
case. We propose that type IIB string theory on $AdS_5 \times H$, with 
$H$ being the horizon manifold of a $D_4$ singularity, is dual to
the superconformal field theory obtained by letting flow 
to the infrared fixed point a large $N$ $\CN =1$ $SU(N)^3$ gauge theory
with matter content \fields\ and then perturbing it with the 
superpotential \sp.

We will test this duality by comparing moduli spaces in both 
theories, finding the string counterpart of the global symmetries 
of the gauge theory and by matching
baryon operators of the gauge theory with wrapped branes 
on $H$. For this we need to know the homology of $H$.
We have seen in the previous section that the space 
$K$ can be obtained, when $a_i \neq 0$, from the intersection 
between the four-fold $Q$ and the hyperplane \CYadd. 
The four-fold $Q$ has three codimension three subspaces of 
singularities. When $a_i \neq 0$ the only singular point 
contained in the hyperplane is the origin of $Q$. The space 
$K$ has thus an isolated singularity at the origin and therefore
its horizon manifold $H$ will be a smooth five-dimensional space.
However when one of the constants $a_i$ is zero, the hyperplane 
\CYadd\ intersects one of the singular subspaces of $Q$. The space we
obtain is then $\tilde K$, with defining equation \spp. 
This space was analyzed in detail in \mope. It contains a complex line 
of $\QZ_2$ singularities. As a result of this, its horizon manifold
$\tilde H$ is singular along an $S^1$. The space $\tilde H$ admits a 
description as $(S^3 \times S^3) / U(1)$, similar to the conifold 
\foot{However contrary to the conifold the $U(1)$ action is not regular, 
i.e. the orbits of the $U(1)$ action do not have constant length. 
This gives rise to the circle of singularities.}. 
It has a non-trivial two-cycle and a non-trivial three-cycle. 
The transversal space to the singular $S^1$ in $\tilde H$ can be 
described locally as $\QC^2 / \QZ^2$, and resolved by replacing the 
singular point by a two-sphere $\Sigma$. This process generates an 
additional three-cycle in $\tilde H$ given by $S^1 \times \Sigma$, 
where $S^1$ is the circle of singularities.  

Since both $\tilde K$ and $K$ can be obtained from the four-fold
$Q$ by just continuously varying the orientation of the hyperplane
\CYadd, and the horizon manifold $H$ of $K$ is non-singular, we could 
interpret $H$ as a deformation of $\tilde H$ that smoothes out the 
singularities. Based on this heuristic argument, we will assume the 
simplest situation that the second and 
third Betti numbers of $H$ are $b_2=b_3=2$. Support for this hypothesis 
will come from obtaining a consistent $AdS/CFT$ correspondence. In 
this way the duality proposed by Maldacena can also be used to learn 
about the topology of singular spaces \milnor, which is in many cases 
an open problem. 

As a first check, we compare the moduli space of string and 
superconformal field theory. With the previous hypothesis for the homology 
groups of $H$, type IIB string theory on $AdS_5 \times H$ possesses a 
(complex) three dimensional moduli space. The moduli parameters are 
the complexified string coupling constant and two additional complex 
parameters coming from integrating the $B$ fields over the two homology 
two-cycles. 
 
The presence of truly marginal deformations of the conformal
theory can be analyzed with the methods of \st. The scale dependence of the 
gauge couplings and coupling constants appearing in the superpotential 
is governed by the following quantities
\eqn\marg{
\eqalign{
A_{g_1} \propto A_{h_{AC}}=1+\gamma_A + \gamma_C,  \;\;\;\;\;\;\;\; 
& A_{h_{A^2}}=1+2\gamma_A, \cr
A_{g_2} \propto A_{h_{BA}}=1+\gamma_B + \gamma_A,  \;\;\;\;\;\;\;\;
& A_{h_{B^2}}=1+2\gamma_B, \cr
A_{g_3} \propto A_{h_{CB}}=1+\gamma_C + \gamma_B,  \;\;\;\;\;\;\;\;
& A_{h_{C^2}}=1+2\gamma_C,}
}
where $\gamma_{X_i}$ are the anomalous dimensions of the fields, 
$X_i=A,B,C$. We have used $\gamma_{X_i}= \gamma_{{\tilde X}_i}$
because charge conjugation is a symmetry of the theory.
The vanishing of all \marg\ only imposes three relations. 
Since we want to preserve the geometrical interpretation of
the gauge theory moduli space, it is crucial that the superpotential 
contains only three independent coupling constants, as in \sp.
We have then six couplings and three relations they must
satisfy. The conformal theory we are considering will thus
have three marginal couplings, as its proposed string dual.  

We now turn to obtain the string counterpart to the global symmetries of 
the gauge theory. It will be convenient
to use the description of $K$ in terms of the variables 
$(x_1,x_3,z,w)$, since they have a direct gauge theory 
interpretation. The space $K$ has a scaling symmetry 
$(x_1, x_3,z,w) \rightarrow (\lambda^2 x_1, \lambda^2 x_3 , 
\lambda^3 z, \lambda^3 w)$, with $\lambda \in \QC^*$.
Restricting $\lambda= e^{i \alpha}$ we obtain a set of $U(1)$
transformations that act on the holomorphic top form
\eqn\top{
\Omega= {d x_1 \wedge d x_3 \wedge d z \over z}
}
by $\Omega \rightarrow e^{4 i \alpha} \Omega$. This is thus an 
R-symmetry in type IIB string theory \foot{We are assuming again that 
$K$ is a non-compact Calabi-Yau three-fold. Then $\Omega = 
\eta \Gamma \eta^T$, with $\eta$ a covariantly constant spinor. Thus
transformations that act on $\Omega$ are R-symmetries.} which 
we can match with
the non-anomalous $U(1)_R$ symmetry of the field theory \kw. 
The space $K$ is also invariant under the transformation $z \rightarrow
e^{i \beta} z$, $w \rightarrow e^{-i \beta}w$. Since the top form 
is invariant under this transformation, it will not be an 
R-symmetry. We can associate it with the last $U(1)$ in Table.1.  
We have argued that $b_3(H)= 2$. Integrating the self-dual 
four-form of type IIB theory on the two non-trivial three-cycles we 
get two $U(1)$ fields on $AdS_5$. According to \gkp\ \holw, gauge 
fields in $AdS$ couple to global symmetry currents on the boundary.
In this way the existence of two three-cycles implies two
global $U(1)$ symmetries in the field theory. We can
identify them with $U(1)_1$ and $U(1)_2$ in Table.1.

For generic $a_i$ parameters in the superpotential the only 
discrete symmetry of the gauge theory is charge conjugation.
However when some of the $a_i$ are equal there are additional 
discrete symmetries. Let us consider $a_1=a_3$. 
In this case, the superpotential is invariant under 
the interchange of the second and third $SU(N)$ gauge factor
together with the following action on the matter fields 
\eqn\perm{
A \leftrightarrow {\tilde C}, \;\;\;\;\; {\tilde A} \leftrightarrow C, 
\;\;\;\;\; B \leftrightarrow {\tilde B}.
}
We denote this transformation by ${\cal P}_{23}$.
In terms of the string dual it corresponds to interchanging
the internal space coordinates $x_1 \leftrightarrow x_3$ and
$z \leftrightarrow w$. The top form \top\ remains invariant 
under this operation in accordance with the fact that 
${\cal P}_{23}$ is not an R-symmetry of the gauge theory.
The composition of charge conjugation with ${\cal P}_{23}$
is also a symmetry of the gauge theory. Under it the second 
and third gauge factor are interchanged and the matter fields 
transform as
\eqn\sl{
\eqalign{
& A \rightarrow C^t, \;\;\;\;\; C \rightarrow A^t, 
\;\;\;\;\; B \rightarrow B^t,     \cr
& {\tilde A} \rightarrow {\tilde C}^t, \;\;\;\;\; {\tilde C} \rightarrow 
{\tilde A}^t, \;\;\;\;\; {\tilde B} \rightarrow {\tilde B}^t. }
}
In the dual string theory this
transformation should correspond to acting with $\omega$, the
center of $SL(2;\QZ)$, and interchanging $x_1 \leftrightarrow 
x_3$. 

String theory compactified on $K$, when $a_1=a_3$, has three
$\QZ_2$ symmetries: {\it i)} the interchange of the $x_1$ and 
$x_3$ coordinates in the defining equation of $K$; 
{\it ii)} the interchange of $z$ and $w$; {\it iii)} the center 
of $SL(2;\QZ)$ of type IIB string theory, $\omega$. We have 
found that the composition of any two of these symmetries 
corresponds to a discrete symmetry of the associated field 
theory. We would like to propose that $\omega$ corresponds 
in field theory terms to the interchange of the second and third 
gauge factor and the action on the matter fields $X_i \rightarrow 
X_i^t$, ${\tilde X}_i \rightarrow {\tilde X}_i^t$, with $X_i=A,B,C$. 
It is important to notice that this transformation maps the 
$SU(N)^3$ gauge theory with matter content \fields\ to an 
equivalent theory where the matter fields transform as
\eqn\fieldsinv{       
\eqalign{       
A=({\bf \bar N},1,{\bf N}), \;\;\;\; B= (1,{\bf N},{\bf \bar N}),  
\;\;\;\; & C =  ({\bf N},{\bf \bar N},1), \cr   
{\tilde A}=({\bf N},1,{\bf \bar N}), \;\;\;\; {\tilde B}=(1,{\bf \bar N},
{\bf N}), \;\;\;\; & {\tilde C}=({\bf \bar N},{\bf N},1).}
} 
Support for this interpretation is the following.
In \u\ \muki, D3-branes at a conifold point were 
mapped into a brane configuration in type IIA, or M-theory, by using 
T-duality. The dual configuration consists of NS5-branes expanding
along $012345$ and $012389$ directions and D4-branes expanding
along $01236$ with the $x_6$ coordinate living on a circle.
In \muki\ it was proposed that $\omega$ of type IIB corresponds,
in the dual M-theory set-up, to the transformation 
$x_6 \rightarrow -x_6$, $x_{10} \rightarrow -x_{10}$. 
We saw in the previous section that our $\CN=1$ gauge 
theory can be derived from an $\CN=2$ theory with the same
matter content by integrating out the adjoint fields. The $\CN=2$
theory can be easily dualized to an elliptic model \wtwo\ with three
parallel NS5-branes (see Fig.1). Let us associate $x_6=0$ to the
position of the NS5-brane originating the fields $B$ and
$\tilde B$. The transformation $x_6 \rightarrow -x_6$, $x_{10}
\rightarrow -x_{10}$ has the
same effect as we have proposed for $\omega$: it interchanges
the second and third gauge factor without interchanging
the matter fields. This brings the initial $\CN=2$ theory 
to an equivalent theory with matter content precisely as in 
\fieldsinv. 
The action of $x_1 \leftrightarrow x_3$, and $z \leftrightarrow w$ 
on the field theory can then be deduced from $\omega$ and \perm, \sl.
\ifig\ads{Type IIA brane configuration dual to D3-branes at a
$\QC^2/\QZ_3$ orbifold.}
{
\epsfxsize=3truein\epsfysize=2.5truein
\epsfbox{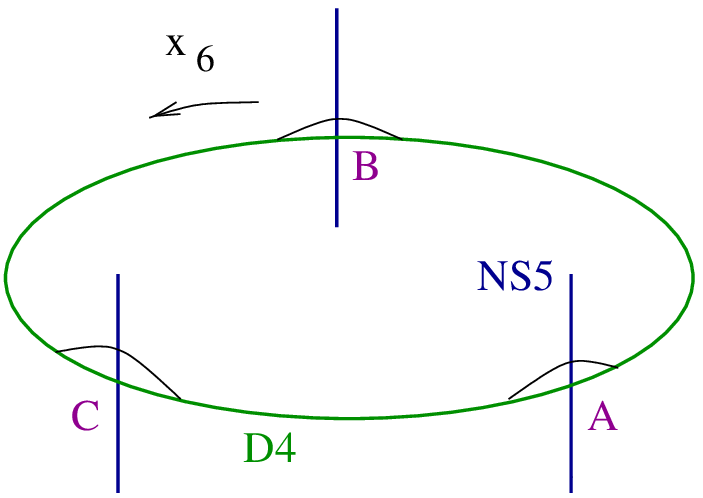}}

Finally, we would like to match baryons of the gauge theory with
D3-branes wrapped on three-cycles of the horizon manifold $H$ \gk\ \w.  
We have conjectured that $H$ has two non-trivial three-cycles. 
The integration of the self-dual four-form of type IIB on these 
cycles gives two $U(1)$ gauge fields in $AdS_5$. We have argued
that they induce on the boundary field theory the global $U(1)_1$ and 
$U(1)_2$ symmetries of Table.1. Let us denote by $C_1$ and $C_2$ the 
three-cycles associated to $U(1)_1$ and $U(1)_2$ respectively. 
D3-branes wrapped around $C_i$ will correspond in the field
theory to baryons charged under $U(1)_i$, $i=1,2$. Therefore 
we can read from the quantum numbers of the matter fields  
how to relate wrapped D3-branes with baryon operators
\eqn\baryons{
B_A \rightarrow \; C_1-\,C_2, \;\;\;\; B_B \rightarrow \; C_2 , \;\;\;\;
B_C \rightarrow \,-\;C_1.
}
The same relation holds for the antibaryons $B_{\tilde X}$,
when instead of D3-branes we wrap anti D3-branes. 

When one of the $a_i$ parameters in the
superpotential is zero $H$ degenerates into the singular space
$\tilde H$, the horizon manifold of $\tilde K$ given by \spp. 
According to our picture,
in this limit a two-cycle and a three-cycle of $H$ shrink to
zero size. We can use this fact to identify the cycles $C_1$ 
and $C_2$. The field theory on D3-branes at 
$\tilde K$ is $\CN=1$ $SU(N)^3$ with matter content  
\fields\ plus a chiral multiplet transforming in the adjoint 
representation of 
one of the gauge groups \mope\ \u. In \mope\ it was argued that
the baryon operator associated with the bifundamental field
which does not couple to the adjoint, corresponds to a D3-brane
wrapped on the shrinked three-cycle. We have shown 
that we can recover this gauge theory from our initial 
one by performing a double limit in which we send one of the 
$a_i$ to zero and the other two to infinity. From that process
we observe that if $a_1$ ($a_2$) ($a_3$) is sent to zero, the 
bifundamentals $A$, $\tilde A$ ($B$, $\tilde B$) ($C$, $\tilde C$) 
do not couple to the extra adjoint. Together with the assignations
\baryons, this implies that $C_1$ is the three-cycle that
shrinks to zero size when $a_3=0$, and $C_2$ is the one which shrinks 
when $a_2=0$. When $a_1=0$, the combination $C_1-C_2$ will shrink 
to zero size.

\newsec{General Case.}

In this section we want to determine what type of singular three-folds
could lead to theories on D3-branes that generalize our previous one. 
In particular, we are interested in obtaining as world-volume theory 
an $\CN=1$ gauge theory with group $SU(N)^k$ and chiral multiplets
\eqn\kfields{
X_i= ({\bf N_i},{\bf \bar N_{i+1}}), \quad 
{\tilde X}_i= ({\bf \bar N_i},{\bf N_{i+1}})
}
for $i=1,..,k$. We will fix again the superpotential by the condition 
that a branch of the moduli space can be interpreted as positions of 
D3-branes in a transversal space.

We will follow the same steps as in section 1. We begin by considering 
a $U(1)^k$ theory with matter content \kfields. We can construct the 
space of solutions to the D-term equations by using gauge invariant 
quantities and moding by the constraints that they are subject to 
\taylor. A minimal set of gauge invariant quantities is
\eqn\kCYfields{
x_i=X_i {\tilde X}_i, \;\;\;\; z=X_1 X_2 \dots X_k , \;\;\;\; 
w={\tilde X}_1 {\tilde X}_2 \dots {\tilde X}_k.
}
These quantities are subject to the relation
\eqn\kCY{
x_1 x_2 \dots x_k = zw,
}
which defines a hypersurface in $\QC^{k+2}$. Let us denote it by $Q$, 
$dim_{\QC} Q=k+1$. The moduli space of the gauge theory will be the 
subspace of $Q$ determined by the F-term equations. The most general
superpotential containing only quartic terms is
\eqn\kspone{
W= \sum_{i=1}^k (a_i x_i^2 + 2 b_i x_i x_{i-1}).
}
The associated F-term equations imply relations expressible in 
terms of the variables $x_i$
\eqn\kfterm{
b_i x_{i-1} + a_i x_i + b_{i+1} x_{i+1} =0,
}
for $i=1,..,k$, with $0 \equiv k$ and $k+1 \equiv 1$. These equations 
define a set of hyperplanes in $\QC^{k+2}$. If we want to interpret the 
moduli space of the gauge theory as positions of a D3-brane, the $2k$ 
parameters appearing in the superpotential cannot be all independent. 
They must be such that only $k-2$ of the $k$ relations \kfterm\ are 
linearly independent. Then the moduli space of the gauge theory will 
be given by the intersection between $Q$ and the $k-2$ hyperplanes 
associated to \kfterm. We will denote again the moduli space by $K$; 
with the previous condition $dim_{\QC} K=3$. 

We can take the equations $i=2,..,k-1$ in \kfterm\ as linearly
independent, and use them to express $x_3,..,x_k$ in terms of $x_1$ 
and $x_2$. Let us define matrices $A_i$ and $B_i$, of dimension 
$(i-2)\times(i-2)$ and $(i-3)\times(i-3)$ respectively, by
\eqn\solft{
\eqalign{
A_i= \pmatrix{a_2 & b_3 &     &        & & \cr
              b_3 & a_3 & b_4 &        & & \cr
                  &     & \ddots &     & & \cr
                  &     &        &     b_{i-1} & a_{i-1}\cr},\;\;\;\; & 
B_i=\pmatrix{a_3 & b_4 &     &        & & \cr
             b_4 & a_4 & b_5 &        & & \cr
                 &     & \ddots &     & & \cr
                 &     &     &        & b_{i-1} & a_{i-1}\cr}.
}}
In terms of $A_i$ and $B_i$, and for generic values of the parameters,
the variables $x_3,..,x_k$ can be written as
\eqn\xi{
x_i = (-1)^i\; {b_2 \; det B_i\; x_1 + det A_i\; x_2 \over 
\prod_{2<j\leq i} b_j},
} 
with $det B_3\equiv 1$. Notice that the parameters $a_1$, $b_1$ and 
$a_k$ do not appear in \xi. We can ensure that the additional two
equations $i=1,k$ in \kfterm\ are solved by \xi\ by setting 
\eqn\param{
a_1= b_2^2 \; {det B_k \over det A_k}, \;\;\; 
b_1= (-1)^{k+1} {\prod_{1<j} b_j \over det A_k}, \;\;\;
a_k= b_k^2 \; {det A_{k-1} \over det A_k}.
}
Thus we are left with $2k-3$ free parameters in the superpotential.
We can now substitute \xi\ in \kCY. In a generic situation, and after 
evident rescalings, we obtain the following defining equation for $K$
\eqn\kCYsix{
x_1 x_2 (x_1 +x_2) \prod_{i=1}^{k-3} (x_1 + \alpha_i x_2)=zw,
}
where $\alpha_i=detA_i /( det A_3 det B_i)$. The space \kCYsix\ is only 
singular at the origin, unless $\alpha_i=0,1,\infty$ for some $i$, or 
$\alpha_i=\alpha_j$ for some $i$ and $j$. 

We can easily generalize the superpotential \kspone\ to that of 
an $SU(N)^k$ gauge theory with matter content \kfields\
\eqn\ksp{
W= \sum_{i=1}^k \left( a_i \Tr(X_i {\tilde X}_i)^2 + 
2 b_i \Tr X_i {\tilde X}_i {\tilde X}_{i-1} X_{i-1} \right),
}
where the parameters $a_i$ and $b_i$ are restricted as in \param.
For the same reasons explained in section 2, this theory will
have a family of vacua reproducing $K^N/S_N$. Notice that for 
particular values of the parameters $a_i$ and $b_i$ the space 
defined by \kCY\ and \kfterm\ reproduce some of the orbifolds of 
the conifold treated in \u, i.e. those of the form $x^n y^m=zw$. 
However we are considering them as a degenerate limit of
a bigger family.

The main criterion that we have used in the previous construction
is that the set of solutions to the D- and F-term equations defines
an space of complex dimension three. This rather naive criterion 
is not enough to guarantee that the resulting space $K$
is a consistent compactification space for string theory. In order 
to analyze this point it is convenient to change coordinates to 
$x \sim x_1+x_2$, $y \sim x_1-x_2$. Then \kCYsix\ can be rewritten as
\eqn\cDV{
x^k - zw + y f(x,y)=0,
}
with $f$ a polynomial function of $x$ and $y$. We observe that
the space $K$ has a hyperplane section, $y=0$, whose defining
equation is that of an $A_{k-1}$ singularity in complex dimension
two. Singular three-folds having a surface of ADE singularities
as a hyperplane section containing the singular point, are special 
cases of Gorenstein canonical singularities \r. Therefore it is
consistent to consider string theory compactified on $K$ \mope.
Using this result we propose that the low-energy theory on $N$ D3-branes 
at the singular point of $K$, given by \kCYsix, is $\CN=1$ $SU(N)^k$
with matter fields \kfields\ subject to the 
superpotential \ksp. $U(1)$ fields living on the world-volume
of the D3-branes are expected to decouple in the infrared limit and
thus we are just considering $SU(N)^k$ as gauge group.

Our $\CN=1$ theory is related to an $\CN=2$ theory with the 
same gauge group and matter content, as it was the case for $k=3$.
Such an $\CN=2$ theory can be derived from 
D3-branes at an $A_{k-1}$ singularity of a complex two-fold. 
We can break $\CN=2$ to $\CN=1$ by giving masses to the adjoint 
fields and integrating them out. When the mass terms are of the
form $\sum m_i (\Tr \phi_i^2 - \Tr \phi_{i+1}^2)$, the superpotential
of the corresponding $\CN=1$ theory is of the form \ksp\
with parameters $a_i(m_j)$ and $b_i(m_j)$ satisfying the restriction
\param. We observe from equation \cDV\ that the space $K$ associated 
to the $\CN=1$ theory knows about the space associated to the parent 
$\CN=2$ theory. In particular, \cDV\ can be viewed as a deformation of 
an $A_{k-1}$ singularity of a two-fold  \foot{
We could have a priori expected that $K$ describes an $A_{k-1}$ isolated 
singularity of a three-fold. However this possibility does not seem to be 
consistent with the $AdS/CFT$ correspondence, since the link sphere
of an $A_{k-1}$ singularity of a three-fold 
is homeomorphic to $S^5$ for $k$ odd \milnor.}. 

There is an important difference between \kCYsix\ and the space
\CYsix\ associated to the case $k=3$. While in \CYsix\ we can
eliminate all the parameters by convenient rescalings, in \kCYsix\
there are $k-3$ parameters that cannot be removed. Equation \kCYsix\ 
describes a singularity of a three-fold whose moduli space of complex 
structures is of dimension $k-3$. It is interesting to notice that
the space \kCYsix\ for $k=4$ and $\alpha \neq 0,1,\infty$, after a linear
change of coordinates, can be rewritten as 
\eqn\xnine{
x^4 + a x^2 y^2 + y^4=zw,
}
with $a$ a complex parameter ($a \neq \pm 2$). This is the standard form for 
one of the isolated unimodular singularities that a three-fold can acquire, 
which is denoted by $X_9$ in \arnold. 

We would like to finish with a short remark on the $AdS/CFT$
correspondence for the general case.
Using the methods of \st, we can obtain the number of marginal
deformations of the $SU(N)^k$ theory. The number of
couplings in the theory is $3k-3$, where $2k-3$ of them are parameters
in the superpotential and $k$ of them are gauge coupling constants.
The vanishing of all beta functions imposes $k$ constraints.
Thus the number of marginal deformations is $2k-3$.
We can consider now the supergravity metric representing $N$ 
3-branes in $K$. We will assume again that there exists is Ricci-flat
a metric in 
a neighborhood of the singular point of $K$ of the form \metrick.  
The decoupling of gravity limit will transform that metric into
$AdS_5 \times H$, with $H$ the horizon manifold of $K$.  
The space $H$ inherits a moduli space of dimension $k-3$ from $K$.
The $AdS/CFT$ correspondence will imply that type IIB string theory on
$AdS_5 \times H$ is dual to the superconformal field theory obtained by
letting flow to the infrared the $\CN=1$ $SU(N)^k$ theory, with $N$ large
and $W=0$, and then perturbing it by the superpotential
\ksp. From an analysis analogous to that of section 2, we 
obtain a consistent duality if $b_2(H)=b_3(H)=k-1$.
In particular, the dual string theory would have also $2k-3$
moduli parameters: the complexified type IIB string coupling constant, 
the $k-3$ moduli of $H$ and $k-1$ complex parameters coming from 
the integration of the $B$ fields on the homology two-cycles.

\vskip1cm

\centerline{\bf Acknowledgments}
\vskip2mm
    
We thank M. Kreuzer, K. Landsteiner, B. Ormsby and especially 
M. Nikbakht-Tehrani for discussions. This work is supported by an 
FWF project under number P13126-TPH.

\listrefs 
\end